\documentclass{epl}

\title{A numerical renormalization group study of laser induced 
freezing} 
\author{Debasish Chaudhuri\thanks{E-mail: \email{debc@bose.res.in}} \and Surajit Sengupta\thanks{E-mail: \email{surajit@bose.res.in}}}
\institute{
Satyendra Nath Bose National Centre for Basic Sciences -
Block-JD, Sector-III, Salt Lake,
Calcutta - 700098.\\
}

\begin{document}

\maketitle

\begin{abstract}
We study the phenomenon of laser induced  freezing, 
within a numerical renormalization scheme which allows explicit  
comparison with a recent defect mediated melting theory.  
Precise values for the `bare' dislocation fugacities and elastic moduli 
of the 2-d hard disk system are obtained from a constrained Monte Carlo
simulation sampling only configurations {\em without} dislocations. These are  
used as inputs to appropriate renormalization flow equations to obtain the  
equilibrium phase diagram which shows excellent 
agreement with earlier simulation results. We show that the flow equations 
need to be correct at least up to third order in defect fugacity to reproduce 
meaningful results. 

\end{abstract}

\section{Introduction}
Re-entrant ``laser-induced'' freezing\cite{chowdhury,wei} 
(RLIF) has 
received considerable attention\cite{jay,supurna,frey,jcdlvo,surajitlif,lifsd,
lifdlvo1,lifdlvo2,cdas1,cdas2,cdas3} in recent times. A static 
interference pattern obtained by two crossed laser beams provides 
an external potential periodic in one dimension (1-d) 
which induces a system of dielectric colloidal particles,
confined in two dimensions (2-d), to freeze.  
Surprisingly, a further increase in potential 
strength causes a {\em reentrant melting}\cite{wei} transition. 
Qualitatively, starting from a liquid phase, the external periodic potential
immediately induces a density modulation, reducing fluctuations which 
leads to solidification. Further increase in potential confines the system
to decoupled 1-d strips. The dimensional reduction now {\em 
increases} fluctuations remelting the system.
The early mean field theories, namely, Landau theory\cite{chowdhury} 
and density functional theory\cite{jay} 
predicted a change from a first order to 
continuous transition with increase in
potential strength and failed to describe the reentrant behavior, 
a conclusion seemingly confirmed by early simulations\cite{jcdlvo}.
Quite generally however, in 2-d, the effect of thermal
fluctuations is substantial and mean field theory may obtain   
qualitatively wrong results especially for fluctuation driven transitions. 
Following the defect mediated disordering approach of Kosterlitz and Thouless
\cite{kt}(KT), Frey, Nelson and Radzihovsky\cite{frey}(FNR) proposed a detailed theory 
for the reentrant transition based on the unbinding of dislocations
with Burger's vector parallel to the line of potential minima.
More accurate simulation 
studies\cite{surajitlif,lifsd,lifdlvo1,lifdlvo2,cdas1,cdas2,cdas3} on 
systems of hard disks\cite{surajitlif}, soft disks\cite{lifsd}, 
DLVO\cite{lifdlvo1, lifdlvo2} etc. confirmed the 
reentrant freezing-melting transition in agreement with experiments\cite{wei}
and FNR theory. A systematic finite size scaling analysis\cite{surajitlif} of 
simulation results for the 2-d hard disk system in a 1-d modulating 
potential showed, in fact, several universal features consistent with the 
predictions of FNR theory. Non universal predictions, namely the phase 
diagram, on the other hand, are difficult to compare because 
the FNR approach (like KT theory) is expressed in terms of the appropriate 
elastic moduli which are notoriously time-consuming to compute near a 
continuous phase transition. Diverging correlation lengths and times near
the phase transition point further complicate an accurate evaluation of the 
non universal predictions of the theory.  
\vskip .2cm

\begin{figure}[t]
\onefigure[width=8.6cm]{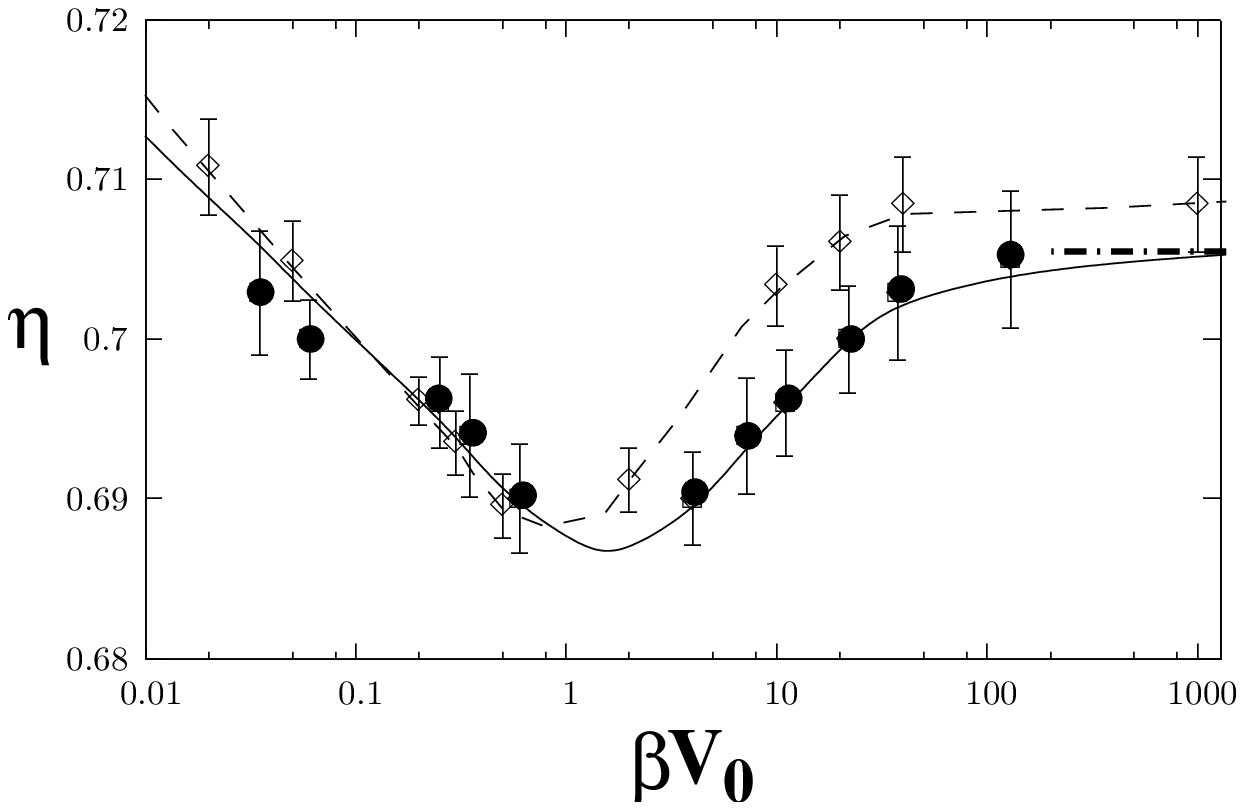}
\caption{The phase diagram of the hard disk system in the presence of a
1-d, commensurate, periodic potential in the packing fraction ($\eta$) -
potential strength ($\beta V_0$) plane. The lines in the figure are a guide
to the eye. The dashed line denotes earlier Monte Carlo simulation
results[7] and the solid line is calculated through our numerical
renormalization group study. The dash-dotted line at $\eta\simeq.705$ 
denotes the
calculated asymptotic phase transition point at $\beta V_0=\infty$.}
\label{phdia}
\end{figure}

\noindent
In this Letter, we calculate the phase diagram of a 2-d hard disk system with 
a modulating potential (see Fig.~\ref{phdia}) following a Monte Carlo 
renormalization approach proposed recently\cite{surajitxy} for the XY-model. 
The twin problems of diverging length and time scales are eliminated by 
simulating a constrained system which {\em does not} undergo a phase 
transition! 
This is achieved by rejecting all Monte Carlo moves which tend to distort 
an unit cell in a way which changes the local 
connectivity\cite{srnb}. The percentage of moves thus rejected is a 
measure of the dislocation fugacity\cite{srnb}. This, together with 
the elastic constants of the dislocation free lattice obtained separately, are 
our inputs (bare values) to the renormalization flow equations\cite{frey}
to find out the melting points and hence the phase diagram. 
Our results (Fig.~\ref{phdia}) clearly show a modulated liquid (ML) 
$\to$ locked floating solid (LFS) $\to$ ML re-entrant transition 
with increase in the amplitude ($V_0$) of the potential. In general, 
we find,  the predictions of FNR theory to 
be valid. The location of the phase transition as evaluated within this 
theory with our inputs, show {\em excellent} agreement  with 
earlier simulations\cite{surajitlif} throughout the $\eta-\beta V_0$ plane
($\beta = 1/k_B T$ with $k_B = $ Boltzmann constant and $T = $ temperature, $\eta=$ packing fraction).
We discuss our calculations in detail below.
\vskip .2cm

\noindent
The bulk system of hard disks where particles $i$ and $j$, in 2-d, 
interact via the potential $V_{ij} = 0$ for $|{\bf r}_{ij}| > {\rm d}$ and 
$V_{ij} = \infty$ for $|{\bf r}_{ij}| \leq {\rm d}$, where ${\rm d}$ is 
the hard disk diameter and ${\bf r}_{ij} = {\bf r}_j - {\bf r}_i$ the 
relative position vector of the particles, is known 
to melt\cite{alzowe,jaster,srnb} from a high 
density triangular lattice to an isotropic liquid with a narrow 
intervening hexatic phase\cite{kthny,jaster,srnb}. 
Simulations\cite{jaster}, experimental\cite{colbook} and 
theoretical\cite{rhyzov} studies of hard 
disks show that for $\eta > .715$ the system exists as a triangular 
lattice which transforms to a liquid below $\eta = .706$. The small 
intervening region contains a hexatic phase predicted by 
the KTHNY theory\cite{kthny} of 2-d melting. 
 Apart from being easily 
accessible to theoretical treatment\cite{hansen-macdonald}, experimental systems
with nearly ``hard'' interactions viz. sterically stabilized 
colloids\cite{colbook} are available.  
The hard disk free energy is entirely entropic in 
origin and the only thermodynamically relevant variable is the number density   
$\rho = N/V$ or the packing fraction $\eta = (\pi/4) \rho {\rm d}^2$. 
\vskip .2cm

\begin{figure}[t]
\onefigure[width=5.0cm]{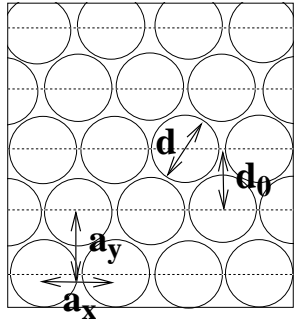}
\caption{This cartoon shows a typical hard disk system. The dashed lines
indicate minima of external modulating potential $\beta V(y)=\beta V_0 \cos(2\pi y/d_0)$.
$a_x$ is the lattice parameter and $a_y$ indicate the average separation 
between two layers along $y$-direction perpendicular to a set of close-packed 
planes. For a perfect triangular lattice $a_y=\sqrt{3}a_x/2$.
The modulating potential is commensurate with the lattice such that
$d_0=a_y$.}
\label{cartoon}
\end{figure}
\vskip .2cm

\noindent
\section{Theory}
In presence of a periodic external potential, the only other energy scale 
present in the system is the relative potential\cite{dielec}
strength $\beta V_0$. A 
cartoon of the system considered for our study is given in Fig.~\ref{cartoon}.
For a solid in presence of a modulating potential $\beta V(y)$
(Fig.~\ref{cartoon})
displacement mode $u_y$ becomes massive, leaving massless $u_x$ modes. After
integrating out the $u_y$ modes the free energy of the LFS 
may be expressed in terms of $u_x$ and $\beta V_0$  
dependent elastic moduli\cite{frey}, namely, the Young's
modulus $K$ and shear modulus $\mu$,
\begin{equation}
 {\cal H}_{el} = \int dx dy \left[ K\left(\frac{\partial u_x}{\partial x}\right)^2+\mu\left(\frac{\partial u_x}{\partial y}\right)^2\right]
\end{equation} 
\vskip .2cm

Similar arguments\cite{frey} show that among the three sets of low 
energy dislocations available in the 2-d triangular lattice, only those 
(type I) with Burger's vector parallel to the line of potential minima survive 
at large $\beta V_0$. Dislocations with Burger's vector pointing along the 
other two possible close-packed directions (type II) in the 2-d triangular 
lattice have larger energies because surrounding atoms are forced to ride 
the crests of the periodic potential\cite{frey}. Within this set of 
assumptions, the system therefore shares the same symmetries as 
the XY model. Indeed, 
a simple rescaling of $x\to\sqrt{\mu}x$ and $y\to\sqrt{K}y$ leads this free
energy to the free energy of the XY-model with spin-wave stiffness 
$K_{xy}=\sqrt{K\mu}a^2/4\pi^2$ and spin angle $\theta=2\pi u_x/a_x$. 
The corresponding theory for phase transitions can then be recast as a  
KT theory\cite{kt} and can be described in the framework of a two parameter 
renormalization flow for the spin-wave stiffness $K(l)$ and the fugacity of 
type I dislocations $y'(l)$,
where $l$ is a measure of length scale as $l=\ln(r/a_x)$, $r$ being the 
size of the system. The flow equations can be expressed in
 terms of $x'=(\pi K_{xy}-2)$ and $y'=4\pi~exp(-\beta E_c)$ where $E_c$ is the
core energy of type I dislocations which can be obtained from the dislocation 
probability\cite{srnb}.
Keeping  upto next to leading order terms in $y'$ 
the renormalization group flow equations\cite{amit,surajitxy} 
are,
\begin{eqnarray}
\frac{dx'}{dl}&=& -y'^2 - y'^2x' \nonumber \\ 
\frac{dy'}{dl}&=& -x'y' + \frac{5}{4}y'^3.
\label{floweq}
\end{eqnarray}
Flows in $l$ generated by the above equations starting from initial or ``bare''
values of $x'$ and $y'$ fall in two categories. If, 
as  $l\to\infty$,  $y'$ diverges, 
the thermodynamic phase is disordered (i.e. ML), while on the other hand
if $y'$ vanishes, it is an ordered phase (a LFS)\cite{frey}. 
The two kinds of flows are demarcated by the {\em separatrix} 
which marks the phase transition point. For the linearized equations the 
separatrix is simply the straight line $y' = x'$, whereas for the full 
non-linear equations one needs to calculate this numerically.   
\vskip .2cm

\noindent
\section{Simulation results and Discussion}
\begin{figure}[t]
\onefigure[width=10cm]{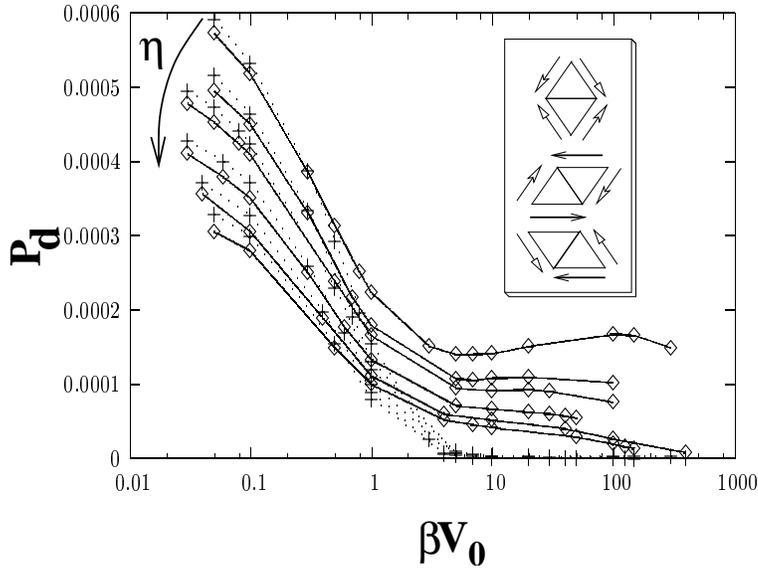}
\caption{Inset shows the unit cell and three possible 
bond-breaking moves for a triangular solid in the orientation shown in 
Fig.~\ref{cartoon}. Arrows show the Burger's vectors for the associated 
dislocations. Dislocations with Burger's vector parallel to the horizontal 
direction are type I dislocations. The probabilities for these moves are 
$P_{m1}$(top), $P_{m2}$ and $P_{m3}$. In the main plot the {\Large $\diamond$} 
symbols 
correspond to $P_{d1}$, the probability for type I dislocations and 
the {\large $+$} symbols to  $(P_{d2}+P_{d3})/2$ the probability for type II 
dislocations
obtained from the $P_{mi}$ (see text)
for various $\eta$ values, arrow denoting the direction of increasing
$\eta$($=.69,.69395,.696,.7,.7029,.705$).
These probabilities are plotted against the  potential strength 
$\beta V_0$. Note that for $\beta V_0 > 1$, the probability for type I
dislocations is larger than that of type II. 
}
\label{dislo}
\end{figure}

We obtain the bare $y'$ and $x'$ from Monte Carlo simulations of
$43\times 50=2150$ hard disks and use them as initial 
values for numerically solving the Eqs.\,(\ref{floweq}). The bare numbers are 
relatively insensitive to system size since our Monte Carlo simulation does not
involve a  diverging correlation length. This is achieved 
as follows\cite{surajitxy,srnb}. We monitor individual random moves of a 
hard disk (after checking for overlaps with neighbors) 
and look for distortions of the neighboring unit cells. If in 
any of these unit cells the length of a next nearest neighbor bond tends to 
become smaller than a nearest neighbor bond, the move is rejected. The 
probabilities of such bond breaking moves are however computed by 
keeping track of the number of such rejected moves. One has to keep 
track of three possible distortions of the unit rhombus (see Fig.~\ref{dislo}
 inset) with measured probabilities $P_{mi}, i = 1,3$. Each of these distortions
involves two dislocation-antidislocation pairs which, we assume, occur 
independently. The probabilities for occurrence of the dislocation pairs 
themselves $P_{di}$ (Fig.~\ref{dislo}) which are proportional to the square 
of the fugacities, can then be computed 
easily eg. $P_{d1}=(P_{m2}+P_{m3}-P_{m1})$.
Finally, we obtain the required dislocation fugacity $y'$
by normalizing $(P_{di})^{1/2}$ to the known\cite{srnb} dislocation 
fugacities for $\beta V_0 = 0$ at any $\eta$. 
The same restricted Monte Carlo simulation can be used to find out the stress 
tensor, and the elastic moduli from the stress-strain curve, 
following the method described by Farago {\em et. al.}\cite{elast}. 
The errors in `bare' elastic moduli are at worst within a percent.
\vskip .2cm

\begin{figure}[t]
\onefigure[width=6cm]{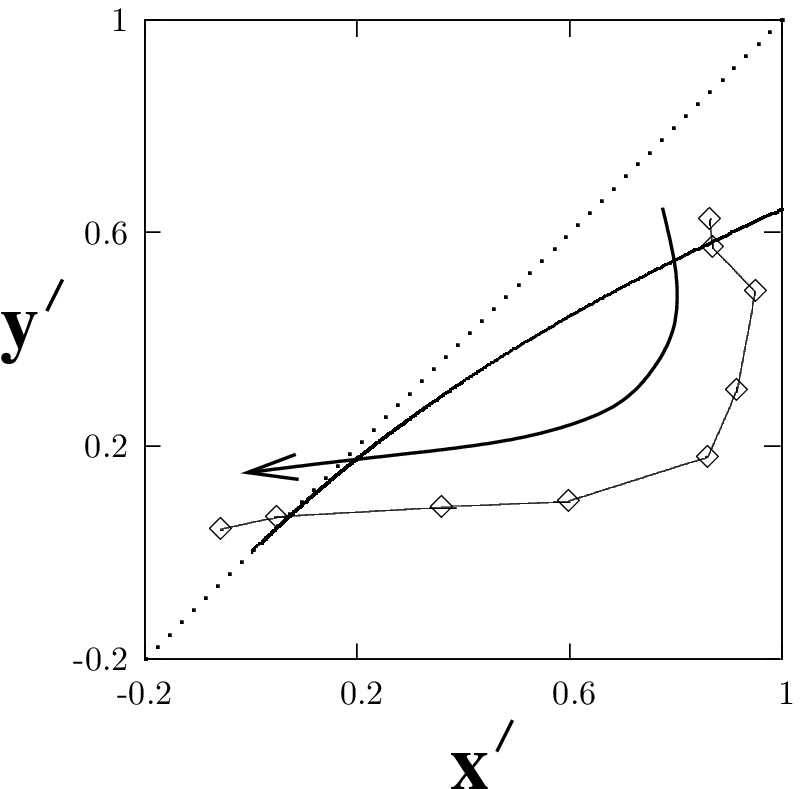}
\caption{
The initial values of $x'$ and $y'$ obtained from the
elastic moduli and dislocation probability
at $\eta=.7029$ plotted in $x'-y'$ plane. The line connecting the points
is a guide to eye. The arrow shows the direction of increase in
$\beta V_0$($=.01, .04, .1, .4, 1, 4, 10, 40, 100$).
The dotted line denotes the seperatrix ($y'=x'$) incorporating upto the
leading order term in KT flow equations. The solid curve is the seperatrix
when next to leading order terms are included. In
$l\to\infty$ limit any initial value below the seperatrix flows to $y'=0$ line
whereas that above the seperatrix flows to $y'\to\infty$. The intersection
points of the line of initial values with
the seperatrix gives the phase transition points. The plot shows a
freezing transition at $\beta V_0=.035$ followed by a melting at $\beta V_0=38$.
}
\label{flow}
\end{figure}

\noindent
In Fig.~\ref{flow} we have plotted the bare values of $x'$ and $y'$
for $\eta=.7029$  
along with the separatrices for the linearized and the non-linear flow 
equations (Eq. \ref{floweq}). The line of initial conditions is seen to 
cross the non-linear separatrix twice (signifying re-entrant behaviour)  
while crossing  the  corresponding linearized separatrix only once at high 
potential strengths. 
The phase diagram (Fig.~\ref{phdia}) is obtained by computing the 
points at which the line of 
initial conditions cut the non-linear separatrix using a simple 
interpolation scheme.  
It is interesting to note that within a linear theory the KT flow 
equations fail to predict a RLIF transition. 

\vskip .2cm
Further, comparing with previous computations\cite{surajitlif} of the phase 
diagram for this system (also shown in Fig.~\ref{phdia})
we find that our results agree at all values of $\eta$ and 
$\beta V_0$. This validates both our method and the quantitative 
predictions of Ref. \cite{frey}.
The effect of higher order terms in 
determining non-universal quantities has been pointed out 
earlier\cite{surajitxy} for the planar rotor model but in the present case 
their inclusion appears to be crucial.   
In view of the fact that the linear flow equation predicts a solid
phase at all but the largest values of $\beta V_0$ we require at least upto
next to leading order corrections in flow to obtain meaningful results.
Even then, we expect our procedure to break down at high packing fractions 
in the $\beta V_0 \to 0$ limit where effects due to the cross-over 
from a KT to a KTHNY\cite{kthny} transition at $\beta V_0 = 0$ become 
significant. Indeed, as is evident from Fig.~\ref{dislo} for $\beta V_0 < 1$ 
the dislocation probabilities of both type I and type II 
dislocations are similar and our process (which involves only type I 
dislocations) fails to produce melting as $\beta V_0 \to 0$ for 
$\eta \geq .705$ --- the solid line in Fig. 1 being an extrapolation from 
our results for smaller $\eta$.
This fact is also supported by  
Ref.\cite{surajitlif} where it was shown that though at $\beta V_0 = 1000$ the 
scaling of susceptibility and order parameter cumulants gave good data 
collapse with values of critical exponents close to FNR predictions, at 
$\beta V_0 = .5$, on the other hand, ordinary critical scaling gave better data collapse than
the KT scaling form, perhaps due to the above mentioned crossover effects. 
In the asymptotic limit of $\beta V_0\to\infty$ the system
freezes above $\eta\sim.705$ which was determined from a separate simulation 
in that limit. This number is very close to the earlier 
value $\eta\sim.71$ quoted in Ref.\cite{surajitlif}. As expected, the 
freezing density in the $\beta V_0 \to \infty$ limit is lower than 
the value without the periodic potential.  
\vskip .2cm

\noindent
\section{Conclusion}
In conclusion, using FNR theory we calculate the phase diagram
for a 2-d system of hard disks under a commensurate modulating potential to
find extremely good agreement with earlier simulated results. 
We show that the re-entrance behavior
is built into the `bare' quantities themselves. To obtain the correct 
phase diagram, however, flow equations
need to be correct at least upto next to leading order terms in the dislocation
fugacity. Our results, especially for small potential strengths, is 
particularly sensitive to these terms. Cross-over effects
from zero potential KTHNY melting transition are also substantial at small
values of the potential.

\vskip 0.2cm
\noindent
\acknowledgments
The authors thank P. Nielaba, W. Strepp, A. Chaudhuri and E. Frey 
for useful discussions; D. C. thanks C.S.I.R., India, for a fellowship. 
Financial support by DST grant SP/S2/M-20/2001 
is gratefully acknowledged.

\vskip 0.5cm

 


\begin{thebibliography}{0}
\bibitem{dielec}
This interaction in colloids is due to polarization of the 
dielectric colloidal particles by the electric field of the laser. Though 
experiments of Refs.\cite{chowdhury,wei} use charged colloids, the 
interaction of hard sphere colloids with laser is similar.
\end{thebibliography}
\end{document}